\documentclass[aps,twocolumn,showpacs,prx,superscriptaddress,floatfix,longbibliography]{revtex4-1}

\usepackage{amsmath,amsthm,amsfonts,amssymb,times,bbm,graphicx}
\usepackage{color}
\usepackage{tabularx}
\usepackage{epsfig}
\usepackage{wasysym}
\usepackage{dcolumn}
\usepackage{bm}
\usepackage{epstopdf}
\usepackage{subfigure}
\usepackage{psfrag}
\usepackage{url}
\usepackage{dsfont}

\newcommand{\bra}[1]{\mbox{$\langle #1 |$}}
\newcommand{\ket}[1]{\mbox{$| #1 \rangle$}}

\newcommand{\bd}[1]{\mbox{$\boldsymbol{#1}$}}
\begin{document}

\title{Hubbard model: Pinning of occupation numbers and role of symmetries}

\author{Christian Schilling}
\email{christian.schilling@physics.ox.ac.uk}
\affiliation{Clarendon Laboratory, University of Oxford, Parks Road, Oxford OX1 3PU, United Kingdom}

\date{\today}

\begin{abstract}
Fermionic natural occupation numbers do not only obey Pauli's exclusion principle, but are even further restricted by so-called generalized Pauli constraints. Such restrictions are particularly relevant whenever they are saturated by given natural occupation numbers $\vec{\lambda}=(\lambda_i)$.
For few-site Hubbard models we explore the occurrence of this pinning effect. By varying the on-site interaction $U$ for the fermions we find  sharp transitions from pinning of $\vec{\lambda}$ to the boundary of the allowed region to nonpinning. We analyze the origin of this phenomenon which turns out be  either a crossing of natural occupation numbers $\lambda_{i}(U), \lambda_{i+1}(U)$ or a crossing of $N$-particle energies. Furthermore, we emphasize the relevance of symmetries for the occurrence of pinning.
Based on recent progress in the field of ultracold atoms our findings suggest an experimental set-up for the realization of the pinning effect.
\end{abstract}


\maketitle

\section{Introduction and concepts}\label{sec:intro}
The Pauli exclusion principle restricts fermionic occupation numbers $n_i$
,
\begin{equation}\label{PauliEx}
0 \leq n_i \leq 1\,.
\end{equation}
From this $1$-particle viewpoint the strong influence of Pauli's exclusion principle for fermionic quantum systems is not surprising. Since it prevents fermions in lattice systems from hopping to a neighboring site which is already occupied by another fermion in the same spin state it can restrict the mobility of fermions significantly.

It has been suspected since the 1970s \cite{Borl1972,Rus2} but was shown only recently \cite{Kly3,Kly2,Altun} that the fermionic exchange statistics
does not only imply Pauli's exclusion principle but leads to even stronger restrictions on fermionic natural occupation numbers. To be more specific, we consider $N$ identical fermions described by pure antisymmetric quantum states $|\Psi\rangle \in\wedge^N[\mathcal{H}_1^{(d)}]$. Here, $\mathcal{H}_1^{(d)}$ is the underyling $d$-dimensional $1$-particle Hilbert space. To each $|\Psi\rangle$ we can assign its \emph{natural occupation numbers} (NON) $\lambda_i$, the eigenvalues of the corresponding $1$-particle reduced density operator
\begin{equation}\label{1RDO}
\hat{\rho}_1 \equiv N\, \mbox{Tr}_{N-1}[|\Psi\rangle \langle\Psi|] =\sum_{k=1}^d\,\lambda_k |k\rangle \langle k|\,.
\end{equation}
$\hat{\rho}_1$ is obtained by tracing out $N-1$ fermions and its eigenstates, $|k\rangle$, are called \emph{natural orbitals}. Note, that the implicit dependence of $\lambda_k$ and $|k\rangle$ on $\hat{\rho}_1$ and $|\Psi\rangle$, respectively, is suppressed. Pauli's exclusion principle (\ref{PauliEx}) can be reformulated as
\begin{equation}\label{PauliExNON}
0\leq \lambda_i\leq 1\,.
\end{equation}
The so-called \emph{generalized Pauli constraints} (GPC) then take the form of linear inequalites \cite{Kly3,Kly2,Altun}
\begin{equation}\label{GPC}
D_j^{(N,d)}(\vec{\lambda}) \equiv \kappa_j^{(0)}+\sum_{i=1}^d\kappa_j^{(i)} \lambda_i\geq 0\,,
\end{equation}
with $\kappa_j^{(i)} \in \mathbb{Z}$, $j=1,\ldots,\nu_{N,d} <\infty$, which give rise to a polytope $\mathcal{P}_{N,d} \subset \mathbb{R}^d$ of allowed vectors $\vec{\lambda}=(\lambda_i)_{i=1}^d$ of decreasingly-ordered NON. The family of GPC implies Pauli's exclusion principle and $\mathcal{P}_{N,d}$ is in particular a proper subset of the `Pauli hypercube' $[0,1]^d$ described by (\ref{PauliExNON}).

A direct significance of GPC was proposed by Klyachko \cite{Kly1,Kly5} for ground states: For some systems the minimization process of the energy $E[\Psi_N]\equiv\bra{\Psi_N}\hat{H}\ket{\Psi_N}$ could get stuck on the boundary of the polytope $\mathcal{P}_{N,d}$ since any further minimization would violate some GPC. In that case, ground state properties would not only rely on the form of the Hamiltonian but also significantly on the structure of GPC.
Such pinning of $\vec{\lambda}$ to the polytope boundary leads to further remarkable consequences.  On the one hand, it potentially restricts the dynamics from the $1$-particle viewpoint since $\vec{\lambda}$ can never leave the polytope. On the other hand, it implies that the corresponding $N$-fermion quantum state
has a significantly simplified structure and also reduced entanglement \cite{Kly1,CSthesis,CSQMath12,CSQuasipinning}.

GPC may have a more indirect physical relevance, as well. For instance, they may lead to improvements in reduced density matrix functional theories as explored in \cite{RDMFT,RDMFT2}, can be used for describing the openness of fermionic quantum systems \cite{MazzOpen} and allow to quantify the influence of the fermionic exchange statistics beyond that of Pauli's exclusion principle \cite{CSQ}.

Based on these ideas, a fundamental question arises: Do there exist fermionic systems exhibiting the pinning-effect? Although Klyachko suggested a mechanism for pinning it would be rather surprising if NON of interacting fermions did \emph{exactly} saturate some of those $1$-particle constraints.
For instance, NON for interacting fermions do never saturate Pauli's exclusion principle, since none of them is ever exactly identical to $1$ or $0$. Further evidence that pinning is unlikely to occur is provided by reduced density matrix functional theory. It seeks a distinguished functional $\mathcal{F}[\hat{\rho}_1]$ whose minimization leads to the energy and the $1$-particle reduced density operator $\hat{\rho}_1$ of the ground state \cite{Gilbert}. Since $\mathcal{F}[\hat{\rho}_1]$  is nonlinear (see e.g.~\cite{DFTbook1}), there is no reason why the minimum of $\mathcal{F}[\hat{\rho}_1]$ should be attained on the polytope boundary.

Given all this evidence against pinning, it is quite surprising that its occurrence was claimed in \cite{Kly1} for the beryllium atom based on numerical data provided in \cite{Nakata2001}. However, by taking additional digits into account (in \cite{Nakata2001} only the first six were shown), this can be disproved \cite{CS2013QChem}. Moreover, in \cite{CS2013} first analytic evidence was found that NON for ground states of interacting fermions do lie indeed close to (but not exactly on) the polytope boundary. The occurrence of this \emph{quasipinning} \cite{CS2013, CSthesis} was also found for several small atoms and molecules \cite{BenavLiQuasi,Mazz14,BenavQuasi2}, provided that the numerical approximations were not too restrictive \footnote{Notice that too elementary approximations lead automatically to pinning. For instance, the Hartree-Fock ansatz implies NON $\vec{\lambda}=(1,\ldots,1,0,\ldots)$ which is a point on the polytope boundary. In a similar way, as correctly noticed in Ref.~\cite{BenavQuasi2}, pinning follows as an artefact of too strong truncations of the $1$-particle Hilbert space in combination with weak correlations.}. All these observations suggest that the occurrence of pinning for non-approximated ground states is highly non-generic, or even impossible.

In this paper, we provide in form of the few-site Hubbard model a first example for a system exhibiting pinning.
Besides its importance for solid state physics the  Hubbard model has gained much relevance in recent years on the microscopic scale, as well. This is due to progress in the field of ultracold fermionic gases which allows to study the crossover from few-fermion to many-fermion physics (see e.g.~\cite{Jochfew2many}, \cite{QOptHub}). In that context, the Hubbard model was realized experimentally very recently\cite{JochimHubb}. The corresponding Hamiltonian (in second quantization) reads
\begin{equation}\label{HamHubbard}
\hat{H} = -t\,\sum_{i=0}^{r-1} \sum_{\sigma}\,\left(c_{i+1\, \sigma}^\dagger\,c_{i \sigma}^{\phantom{\dagger}}\,+\mbox{h.c.}\right)\,\,+\,\,U\,\sum_{i=0}^{r-1}\,\hat{n}_{i\uparrow}\,\hat{n}_{i\downarrow}
\end{equation}
where $c_{i\sigma}^\dagger$ and $c_{i\sigma}$ are the fermionic creation and annihilation operators for a spin-$\frac{1}{2}$ fermion on the $i$-th lattice site with spin $\sigma = \uparrow,\downarrow$ with respect to the $z$-axis and $\hat{n}_{i \sigma}\,\equiv c_{i\sigma}^\dagger\,c_{i\sigma}$.

The paper is organized as follows. In Sec.~\ref{sec:symmetries} we recall the symmetries of (\ref{HamHubbard}) and introduce a symmetry-adapted basis of $N$-fermion quantum states. The main results are presented in Sec.~\ref{sec:BD}. There, we analytically determine the eigenstates of (\ref{HamHubbard}) for three fermions on three lattice sites and explore possible pinning. This is extended by an exact numerical approach for the next larger settings involving more fermions and/or  more lattice sites in Sec.~\ref{sec:larger}. In Sec.~\ref{sec:struct}, we explain the role of symmetries for the occurrence of pinning. By building on very recent progress in the field of ultracold gases we propose in Sec.~\ref{sec:exp} two ideas for experimental realization of pinning. Sec.~\ref{sec:concl} provides a short summary and a conclusion.

\section{Symmetries}\label{sec:symmetries}
In this section we present the most elementary symmetries of the Hubbard model (\ref{HamHubbard}) and introduce a basis of symmetry-adapted quantum states.

Since the symmetries of the Hubbard model are well-known (see e.g.~Ref.~\cite{Fradkin}), they are just listed in order to keep our paper self-contained.
First of all, since the Hamiltonian $\hat{H}$ commutes with the total fermion number operator $\hat{N} = \sum_{i=0}^{r-1} \sum_{\sigma} \hat{n}_{i,\sigma}$
, $[\hat{H},\hat{N}]\,=\,0$, the total particle number $N$ is conserved and we restrict to fixed fermion numbers $N$. By employing first quantization in the following $\hat{H}$ is then restricted to the corresponding $N$-fermion Hilbert space $\mathcal{H}_N \equiv \wedge^N[\mathcal{H}_1]$ of antisymmetric quantum states, where the $2r$-dimensional $1$-particle Hilbert space has the substructure $\mathcal{H}_1=\mathcal{H}_1^{(l)}\otimes\mathcal{H}_1^{(s)}$. Here, $\mathcal{H}_1^{(l)} \cong \mathbb{C}^r$ describes the spatial ($l$) degrees of freedom and $\mathcal{H}_1^{(s)} \cong \mathbb{C}^2$ is the spin Hilbert space of a single fermion.

We denote the total spin vector operator by $\hat{\vec{S}}$, its $z$-component by $\hat{S}_z$ and let $\hat{T}$ be the translation operator which translates each of the $N$ fermions from its lattice sites $i$ to the next site $i+1$.
Then, the symmetries of $\hat{H}$ are described by the following  relations (see e.g.~\cite{Fradkin})
\begin{equation}\label{HamHubbardSym}
[\hat{H},\hat{\vec{S}}^{\,2}]\,=\,0\,\,,\,\,\,[\hat{H},\hat{S}_z]\,=\,0\,\,,\,\,\,\,[\hat{H},\hat{T}]\,=\,0.
\end{equation}
The operators $\hat{\vec{S}}^{\,2}, \hat{S}_z$ and $\hat{T}$ also commute with each other. Consequently, the total spin quantum number $S$, the magnetic spin quantum number $M$ and the total Bloch number (wave number) $K$ are good quantum numbers
\footnote{The Bloch number $K$ for the $1$-dimensional lattice with $r$ sites is given in units of $\frac{2 \pi}{a r}$ ($a$ is the lattice constant) and is restricted to the first Brillouin zone, i.e.~$K=0,1,\ldots,r-1$ and the corresponding $\hat{T}$-eigenstates $\protect{|\Psi_K\rangle}$ fulfill $\hat{T} \protect{|\Psi_K\rangle} = e^{\frac{2 \pi}{r}i K} \protect{|\Psi_K\rangle}$}. The Hamiltonian (\ref{HamHubbard}) is block diagonal w.r.t.~those symmetries and the total Hilbert space splits according to
\begin{equation}
\mathcal{H}_N\equiv \wedge^N[\mathcal{H}_1^{(l)}\otimes \mathcal{H}_1^{(s)}]\, =\bigoplus_{S=S_-}^{S_+}\,
\bigoplus_{M=-S}^S\,\bigoplus_{K=0}^{r-1}\,
\mathcal{H}_{S,M,K}\,.
\end{equation}
By setting $\hbar \equiv 1$ the maximal total spin is given by $S_+ = \frac{N}{2}$ and the minimal total spin $S_-$ by $0$ for $N$ even and $\frac{1}{2}$ for $N$ odd.

Since we will focus below on the $1$-fermion picture it is worth noticing that the $1$-particle reduced density operator $\hat{\rho}_1$ inherits symmetries from the corresponding $N$-fermion quantum states $|\Psi\rangle$ \cite{Dav}. Whenever $|\Psi\rangle$ has a symmetry, $\hat{F}_1^{\otimes^N}|\Psi\rangle = e^{i \varphi}|\Psi\rangle$, where $\hat{F}_1$ is a $1$-fermion operator acting on $\mathcal{H}_1$, the corresponding $\hat{\rho}_1$ inherits the same symmetry, $[\hat{\rho}_1,\hat{F}_1]=0$ and is therefore block-diagonal w.r.t.~the $\hat{F}_1$-eigenspaces.

For the Hubbard model with periodic boundary conditions, the lattice translation $\hat{T}=\hat{T}_1^{\otimes^N}$ defines such a symmetry, where $\hat{T}_1$ is the $1$-fermion translation operator. Another symmetry is generated by the $z$-component of the fermion spin. Consequently, the natural orbitals arising from any symmetry-adapted state $|\Psi\rangle$, $|\Psi\rangle \in \mathcal{H}_{S,M,K}$, are given by $|k\sigma\rangle\equiv |k\rangle \otimes |\sigma\rangle$. Here $|k\rangle \in \mathcal{H}_1^{(l)}$, $k=0,1,...,r-1$ is the corresponding Bloch state, $\hat{T}_1 |k\rangle = e^{\frac{2\pi}{r}i k} |k\rangle$, and $|\sigma\rangle \in \mathcal{H}_1^{(s)}$ , $\sigma=\uparrow,\downarrow$, is the corresponding spin state.
Moreover, since $\hat{\rho}_1$ is diagonal w.r.t.~the NO, we can easily calculate the NON.
They are given by the diagonal elements $\langle k \sigma|\hat{\rho}_1|k \sigma\rangle$.

The basis $\{|k\sigma\rangle\}$ for $\mathcal{H}_1$ induces a basis for the $N$-fermion Hilbert space $\mathcal{H}_N$, the $N$-fermion Slater determinants \begin{equation}\label{SlaterBasis}
|k_1\sigma_1,\ldots,k_N\sigma_N\rangle \equiv \mathcal{A}_N \left[|k_1\sigma_1\rangle\otimes \ldots \otimes|k_N\sigma_N\rangle\right]\,,
\end{equation}
where $\mathcal{A}_N$ is the $N$-particle antisymmetrizing operator and the ordering of the $1$-particle basis states is given by $|0\uparrow\rangle$, $|0\downarrow\rangle, |1\uparrow\rangle,\ldots, |r-1\downarrow\rangle$.

\section{Three fermions on three sites}\label{sec:BD}
In this section we study the Hubbard model (\ref{HamHubbard}) for three fermions on three lattice sites with periodic boundary conditions
\footnote{We do not consider the case of $N=2$ fermions since the corresponding GPC take the form of equalities (see e.g.~\cite{CSQuasipinning}) rather than inequalities and exploring possible pinning is therefore meaningless.}.
We analytically diagonalize the Hamiltonian and explore whether its eigenstates exhibit the pinning or quasipinning effect.

\subsection{Analytic diagonalization of the Hamiltonian}\label{sec:diagH}
To diagonalize (\ref{HamHubbard}) on the corresponding $\binom{6}{3}=20$-dimensional $3$-fermion Hilbert space we first block-diagonalize $\hat{H}$ w.r.t.~to its symmetries which were mentioned in Sec.~\ref{sec:symmetries}.
The total spin quantum number $S$ can take the two values $S=\frac{3}{2}, \frac{1}{2}$. We first split the total Hilbert space w.r.t.~those two values for $S$ and then continue with the magnetic quantum number $M$:
\begin{itemize}
\item $S=\frac{3}{2}$. The only state with maximal magnetic quantum number $M=\frac{3}{2}$ is the  symmetry-adapted state $|0\hspace{-0.1cm}\uparrow,1\hspace{-0.1cm}\uparrow,2\hspace{-0.1cm}\uparrow\rangle$. It has the Bloch number $K=0+1+2 \,(mod\,3)=0$. By successively applying the lowering ladder operator to that state we find the remaining three symmetry-adapted states of the corresponding quadruplet. They read $\frac{1}{\sqrt{3}}\left[|0\hspace{-0.1cm}\downarrow,1\hspace{-0.1cm}\uparrow,2\hspace{-0.1cm}\uparrow\rangle
    +|0\hspace{-0.1cm}\uparrow,1\hspace{-0.1cm}\downarrow,2\hspace{-0.1cm}\uparrow\rangle+
    |0\hspace{-0.1cm}\uparrow,1\hspace{-0.1cm}\uparrow,2\hspace{-0.1cm}\downarrow\rangle\right]$, $\frac{1}{\sqrt{3}}\left[|0\hspace{-0.1cm}\downarrow,1\hspace{-0.1cm}\downarrow,2\hspace{-0.1cm}\uparrow\rangle
    +|0\hspace{-0.1cm}\downarrow,1\hspace{-0.1cm}\uparrow,2\hspace{-0.1cm}\downarrow\rangle+
    |0\hspace{-0.1cm}\uparrow,1\hspace{-0.1cm}\downarrow,2\hspace{-0.1cm}\downarrow\rangle\right]$ and \\ $|0\hspace{-0.1cm}\downarrow,1\hspace{-0.1cm}\downarrow,2\hspace{-0.1cm}\downarrow\rangle$.
As an elementary exercise one verifies that (\ref{HamHubbard}) restricted to that $4$-dimensional subspace is given by $\hat{H}|_{S=\frac{3}{2}}= 0$.
\item $S=\frac{1}{2}$. The corresponding subspace is $16$-dimensional. Since the Hamiltonian is also invariant under simultaneous flips of all spins w.r.t.~the $z$-axis, we restrict without loss of generality to $M>0$, i.e.~$M=\frac{1}{2}$.
    \begin{itemize}
    \item $M=\frac{1}{2}$. In this $8$-dimensional subspace we find two states with $K=0$, \
    $\frac{1}{\sqrt{3}}\big[|0\hspace{-0.1cm}\downarrow,1\hspace{-0.1cm}\uparrow,2\hspace{-0.1cm}\uparrow\rangle$
    $+\,e^{\pm\frac{2 \pi}{3} i}\,|0\hspace{-0.1cm}\uparrow,1\hspace{-0.1cm}\downarrow,2\hspace{-0.1cm}\uparrow\rangle
    +e^{\mp\frac{2 \pi}{3} i}\,|0\hspace{-0.1cm}\uparrow,1\hspace{-0.1cm}\uparrow,2\hspace{-0.1cm}\downarrow\rangle\big]$,
and the Hamiltonian restricted to that $2$-dimensional subspace, $\mathcal{H}_{\frac{1}{2},\frac{1}{2},0}$, is given by $u\,\mathbf{1}_2$.
The remaining subspaces are $\mathcal{H}_{\frac{1}{2},\frac{1}{2},1}$ and $\mathcal{H}_{\frac{1}{2},\frac{1}{2},2}$. However, due to the additional invariance of $\hat{H}$ under inversion in the reciprocal lattice (recall that $K=2$ equals $K=-1$), we can restrict to the case $K=1$. This subspace is spanned by the three states $|0\hspace{-0.1cm}\uparrow,0\hspace{-0.1cm}\downarrow,1\hspace{-0.1cm}\uparrow\rangle$, $|1\hspace{-0.1cm}\uparrow,1\hspace{-0.1cm}\downarrow,2\hspace{-0.1cm}\uparrow\rangle$ and $|0\hspace{-0.1cm}\uparrow,2\hspace{-0.1cm}\uparrow,2\hspace{-0.1cm}\downarrow\rangle$ and $\hat{H}$ represented w.r.t.~to those three states takes the form
\begin{equation}\label{eigenvalueQ}
\qquad\hat{H}|_{\mathcal{H}_{\frac{1}{2},\frac{1}{2},1}}=\left(
\begin{array}{ccc}
 \frac{2 u}{3}-3 & -\frac{u}{3} & -\frac{u}{3} \\
 -\frac{u}{3} & \frac{2 u}{3}+3 & -\frac{u}{3} \\
 -\frac{u}{3} & -\frac{u}{3} & \frac{2 u}{3}
\end{array}
\right)\,.
\end{equation}
The diagonalization of this $3\times3$-matrix leads to a cubic equation which is presented and solved in Appendix \ref{app:cubicEigenProb}.
\item $M=-\frac{1}{2}$. Results are identical to the case $M=\frac{1}{2}$ up to a simultaneous flip of all three spins w.r.t.~the $z$-axis.
\end{itemize}
\end{itemize}
This completes the diagonalization of the Hubbard Hamiltonian (\ref{HamHubbard}).

\begin{figure}[h]
\centering
	\includegraphics[width=6.0cm]{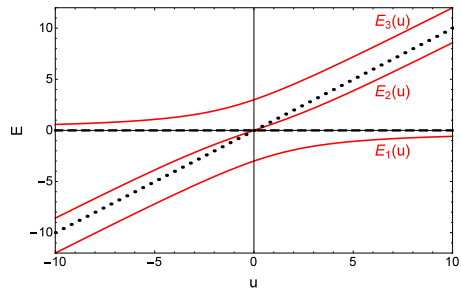}
\caption{Energy spectrum for the 3-site Hubbard model for three spin-$\frac{1}{2}$ fermions. Each eigenvalue has multiplicity four. The dashed black line (horizontal line) corresponds to $S=\frac{3}{2}$.
The case $S=\frac{1}{2}$ yields the black straight dotted line ($K=0$) and the three red solid lines correspond to the solution for the nontrivial subspace problem (\ref{eigenvalueQ}) for $K=\pm1$.}
	\label{fig:energyplots}
\end{figure}

The energy spectrum of $\hat{H}$ is shown in Fig.~\ref{fig:energyplots}.
The three solid lines $E_1(u),E_2(u)$ and $E_3(u)$ are those arising from the eigenvalue problem for (\ref{eigenvalueQ}),
\begin{equation}\label{eigenvalueEq}
\hat{H}|_{\mathcal{H}_{\frac{1}{2},\frac{1}{2},1}} |\Psi_j(u)\rangle = E_j(u)\, |\Psi_j(u)\rangle\,,\,\,j=1,2,3\,.
\end{equation}
According to the symmetries each $E_j(u)$ has multiplicity four (quantum numbers $S = \frac{1}{2}$, $M = \pm \frac{1}{2}$ and $K=\pm 1$). For the four eigenstates with $S=\frac{3}{2}$ we find the eigenvalue $E(u) \equiv 0$ (dashed line) and for those with $S=\frac{1}{2}$, $M= \pm \frac{1}{2}$ and $K=0$, $E(u)=u$ (dotted line).

\subsection{Pinning analysis}\label{sec:pinning}
In this section we investigate whether the energy eigenstates of the Hubbard Hamiltonian are showing quasipinning or pinning. Due to their particular relevance for experiments we first consider symmetry-adapted states. At the end of this section we also study coherent superposition of degenerate eigenstates.

The corresponding GPC (cf.~Eq.~(\ref{GPC})) for the setting of $N=3$ fermions and a $6$-dimensional $1$-particle Hilbert space are given by \cite{Borl1972}
\begin{eqnarray}\label{GPC36}
&&\lambda_1+\lambda_6 = \lambda_2+\lambda_5 =\lambda_3+\lambda_4 =1 \nonumber \\
&&D^{(3,6)}(\vec{\lambda}) \equiv \lambda_5+\lambda_6-\lambda_4 \geq 0\,.
\end{eqnarray}
The first three GPC lead to strong structural implications for the corresponding $3$-fermion quantum state (see e.g.~\cite{CSQuasipinning}). Since they are always saturated it only makes sense to explore a possible saturation of the fourth constraint, which takes the form of a proper inequality.

First, we study the four eigenstates with $S=\frac{3}{2}$. Those two with a minimal/maximal $M$ are \emph{single} Slater determinants and consequently their NON equal the Hartree-Fock point $\vec{\lambda}_{HF} = (1,1,1,0,0,0)$. Since this vector even saturates the weaker Pauli exclusion principle constraint (\ref{PauliEx}) it is also trivially pinned to the polytope boundary. The other two eigenstates, those with $M=\pm \frac{1}{2}$, have NON $\frac{1}{3}(2,2,2,1,1,1)$ which is not pinned. This example also demonstrates that NON of degenerate energy eigenstates can be quite different.

\begin{figure}[h]
\centering
\includegraphics[scale=0.45]{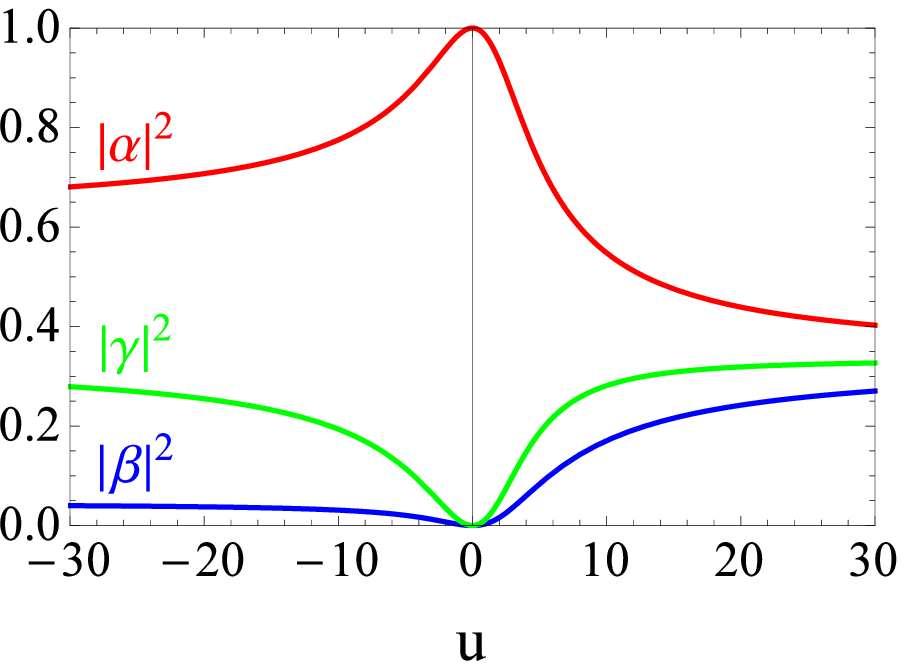}\hspace{0.05cm}
\includegraphics[scale=0.45]{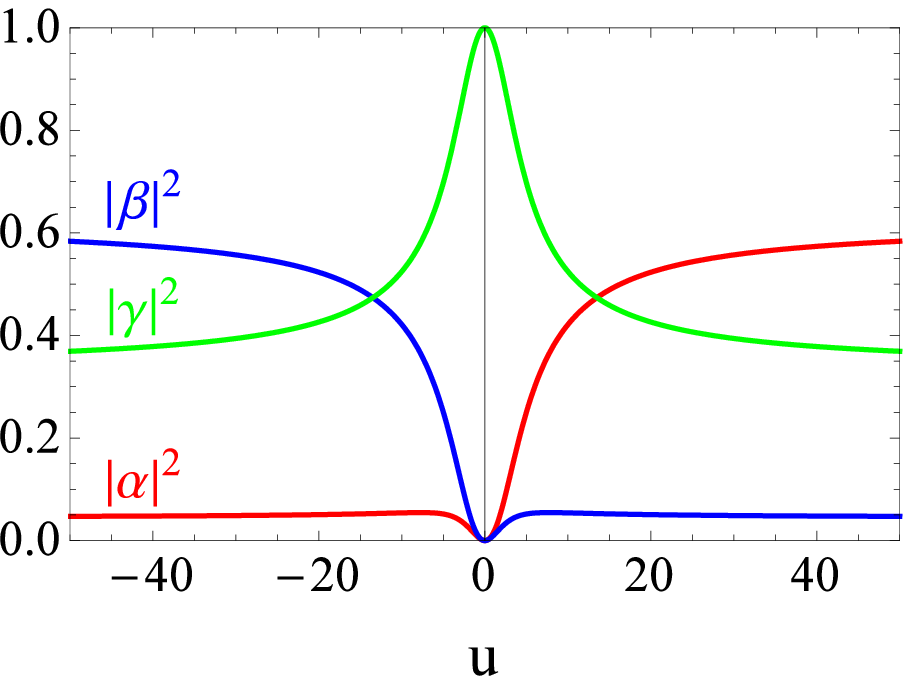}
\caption{The absolute squares of the coefficients $\alpha(u)$ (red), $\beta(u)$ (blue) and $\gamma(u)$ (green) for the eigenstates following from (\ref{eigenvalueEq}). Those  for the ground state $|\Psi_1(u)\rangle$ (left) are identical to those for the highest energy state $|\Psi_3(u)\rangle$ (not shown) up to a reflection w.r.t.~the vertical axis. On the right the corresponding result for the excited state $|\Psi_2(u)\rangle$ is shown.}
\label{fig:HubGScoef}
\end{figure}

Now, we continue with the $S=\frac{1}{2}$-eigenstates. Those four with $K=0$ also do not depend on $u$ and their NON are given again by $\frac{1}{3}(2,2,2,1,1,1)$.
More interesting are the remaining three eigenstates (with multiplicity four) arising from (\ref{eigenvalueQ}) and (\ref{eigenvalueEq}), since they do depend on $u$.
These states have the general form
\begin{eqnarray}\label{gsHubbard1}
|\Psi(u)\rangle &=& \alpha(u)\, |0\hspace{-0.1cm}\uparrow,0\hspace{-0.1cm}\downarrow,1\hspace{-0.1cm}\uparrow\rangle \,+
\beta(u)\, |1\hspace{-0.1cm}\uparrow,1\hspace{-0.1cm}\downarrow,2\hspace{-0.1cm}\uparrow\rangle \nonumber \\
&+& \gamma(u)\, |0\hspace{-0.1cm}\uparrow,2\hspace{-0.1cm}\uparrow,2\hspace{-0.1cm}\downarrow\rangle \,.
\end{eqnarray}
The three coefficients $\alpha(u), \beta(u)$ and $\gamma(u)$ are calculated in Appendix \ref{app:cubicEigenProb}
and their absolute squares are presented in Fig.~\ref{fig:HubGScoef} for the three non-trivial eigenstates following from (\ref{eigenvalueEq}).

The corresponding $1$-particle density operator represented w.r.t.~to the $1$-particle states $|k\sigma\rangle$ is diagonal as explained in Sec.~\ref{sec:symmetries}. In particular, this means that the two different spin-blocks are decoupled
\footnote{This holds in general for all $\hat{S}_z$-symmetry-adapted $N$-fermion quantum states
$\protect{|\Psi\rangle} \in \wedge^N[\mathcal{H}_1^{(l)}\otimes \mathcal{H}_1^{(s)}]$}, i.e.~
\begin{equation}\label{1rdoBlocks}
\hat{\rho}_1 = \hat{\rho}_1^{\uparrow} \oplus \hat{\rho}_1^{\downarrow}\,.
\end{equation}
Since $\hat{\rho}_1$ is normalized to $N=3$ and needs to reproduce the $S_z$-spin expectation value $M=\frac{1}{2}$ the blocks have the specific normalizations $\mbox{Tr}[\hat{\rho}_1^{\uparrow}]=2$ and
$\mbox{Tr}[\hat{\rho}_1^{\downarrow}]=1$.

For the eigenvalues $n_k^{\uparrow/\downarrow}$, $k=0,1,2$ of $\hat{\rho}_1^{\uparrow/\downarrow}$ corresponding to the natural orbitals $|k,\uparrow/\downarrow\rangle$ we find
\begin{eqnarray}\label{NONsHubbard1}
n_0^\uparrow=|\alpha(u)|^2+|\gamma(u)|^2\,, && n_0^\downarrow=|\alpha(u)|^2 \nonumber \\
n_1^\uparrow=|\alpha(u)|^2+|\beta(u)|^2\,,  && n_1^\downarrow=|\beta(u)|^2 \nonumber \\
n_2^\uparrow=|\beta(u)|^2+|\gamma(u)|^2\,,  && n_2^\downarrow=|\gamma(u)|^2\,.
\end{eqnarray}
The structure of this spectrum -- three eigenvalues are given by the weights $|\alpha|^2,|\beta|^2$ and $|\gamma|^2$ and the other three by sums of two of them -- is a consequence of the symmetries, only. The concrete form (beyond the symmetries) of the Hubbard Hamiltonian affects only the weights $|\alpha(u)|^2,|\beta(u)|^2$ and $|\gamma(u)|^2$.

Furthermore, it is an elementary exercise \cite{CSthesis} to show that the normalization of $\hat{\rho}_1^{\uparrow}$ and $\hat{\rho}_1^{\downarrow}$ together with the first three GPC (\ref{GPC36}) implies that the largest two NON, $\lambda_1$ and $\lambda_2$, belong to the block $\hat{\rho}_1^{\uparrow}$, the smallest two NON, $\lambda_5$ and $\lambda_6$, belong to the block $\hat{\rho}_1^{\downarrow}$ and the smallest eigenvalue $n_{min}^\uparrow $ of $\hat{\rho}_1^{\uparrow}$ and the largest eigenvalue $n_{max}^\downarrow$ of $\hat{\rho}_1^{\downarrow}$ fulfill
\begin{equation}\label{NON2cases}
n_{min}^\uparrow+n_{max}^\downarrow=1\,.
\end{equation}

Eq.~(\ref{NON2cases}) simplifies the pinning analysis and we can distinguish two cases
\begin{itemize}
\item $n_{max}^\downarrow \geq n_{min}^\uparrow$: The fourth-largest NON $\lambda_4$ is therefore given by $n_{min}^\uparrow$ and for the second line of Eq.~(\ref{GPC36})
we find
\begin{eqnarray}
D^{(3,6)}(\vec{\lambda}) &=& \lambda_5+\lambda_6-\lambda_4 \nonumber \\
&=& (1-n_{max}^\downarrow)-n_{min}^\uparrow= 0\,,
\end{eqnarray}
where we used that $\lambda_5$ and $\lambda_6$ both belong to $\hat{\rho}_1^\downarrow$, $\mbox{Tr}[\hat{\rho}_1^\downarrow]=\lambda_5+\lambda_6+n_{max}^\downarrow=1$ and Eq.~(\ref{NON2cases}).
\item $n_{max}^\downarrow < n_{min}^\uparrow$. In a similar way we find for this case
\begin{eqnarray}
D^{(3,6)}(\vec{\lambda})&=&  (1-n_{max}^\downarrow)-n_{max}^\downarrow \nonumber \\
&=& n_{min}^\uparrow- n_{max}^\downarrow>0\,.
\end{eqnarray}
\end{itemize}
\begin{figure}[h]
\centering
\includegraphics[scale=0.45]{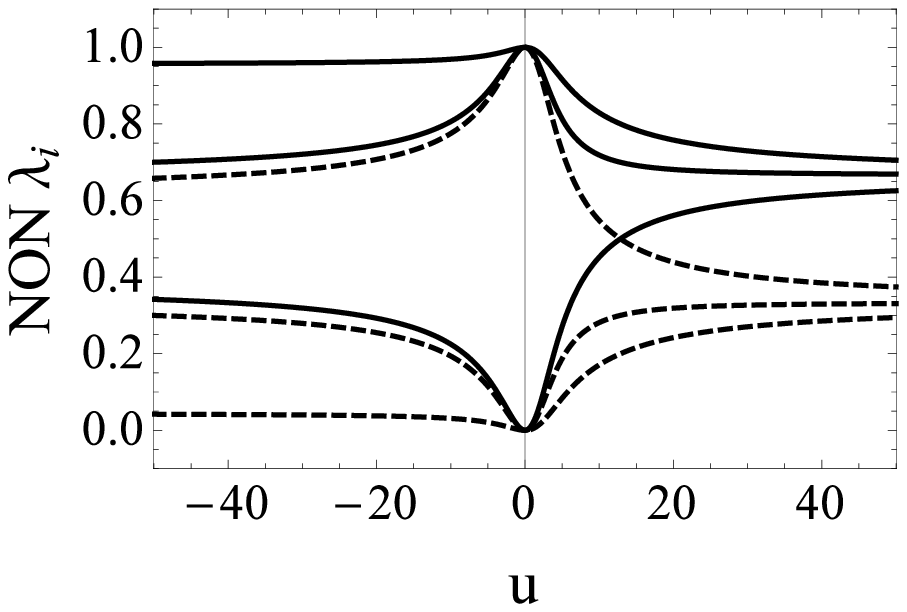}\hspace{0.05cm}
\includegraphics[scale=0.45]{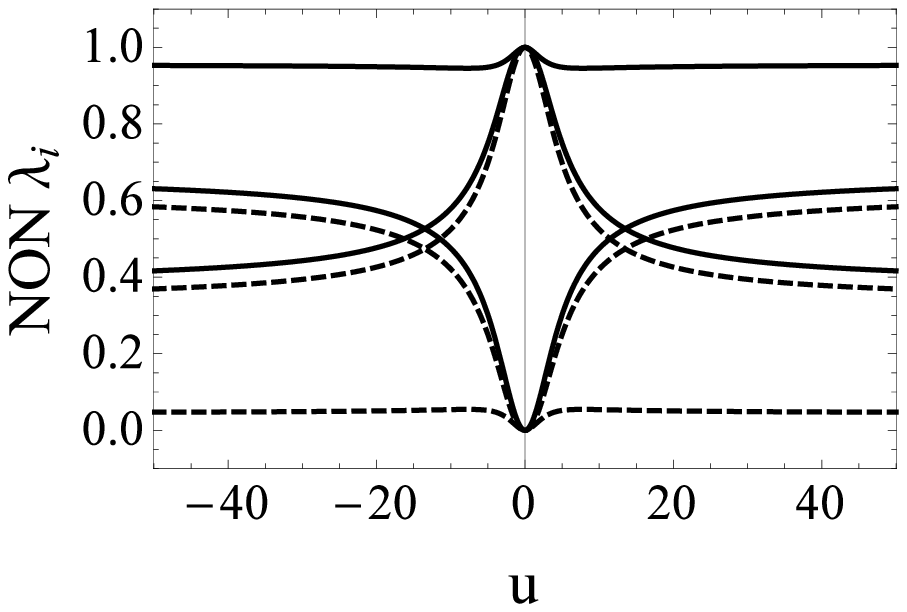}
\caption{u-dependence of the NON $\lambda_i(u)$ for the ground state $|\Psi_1(u)\rangle$ (left panel) and for the excited eigenstate $|\Psi_2(u)\rangle$ (right panel). The NON for the ground state are identical to those for the highest energy state $|\Psi_3(u)\rangle$ (not shown) up to a reflection w.r.t.~the vertical axis. The dashed lines are the NON corresponding to spin $\downarrow$  and the solid lines  correspond to $\uparrow$.}
\label{fig:NON}
\end{figure}
This means, depending on the ratio of $n_{max}^\downarrow$ and $n_{min}^\uparrow$, we either have pinning or nonpinning.
Whether the eigenstates (\ref{gsHubbard1}) are pinned or not is illustrated in Fig.~\ref{fig:NON}.  It shows the six NON
for the relevant on-site interaction regime. The three eigenvalues of the block $\hat{\rho}_1^{\uparrow}$ are shown as solid lines and those for $\hat{\rho}_1^{\downarrow}$ as dashed lines. Pinning is given for those $u$-regimes where $n_{max}^\downarrow \geq n_{min}^\uparrow$, i.e.~whenever the highest
dashed line is above the lowest solid line. E.g.~for the ground state $|\Psi_1(u)\rangle$ presented on the left side of Fig.~\ref{fig:NON} this is
the case for $u \leq  u_0 \approx12.86$. At that `critical' point, $u_0$, there is a transition from pinning to nonpinning according to the crossing of  $n_{max}^\downarrow(u)$ and $n_{min}^\uparrow(u)$
For the excited state $|\Psi_2(u)\rangle$ presented on the right side we can expect an even more complex behavior, since the relevant NON curves do cross several times.

\begin{figure}[h]
\centering
\includegraphics[scale=0.70]{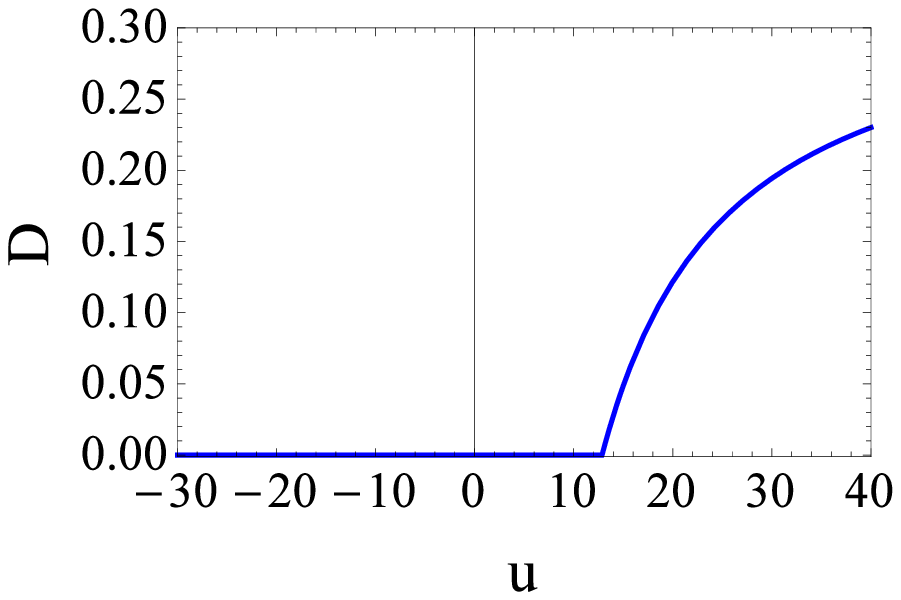}
\includegraphics[scale=0.70]{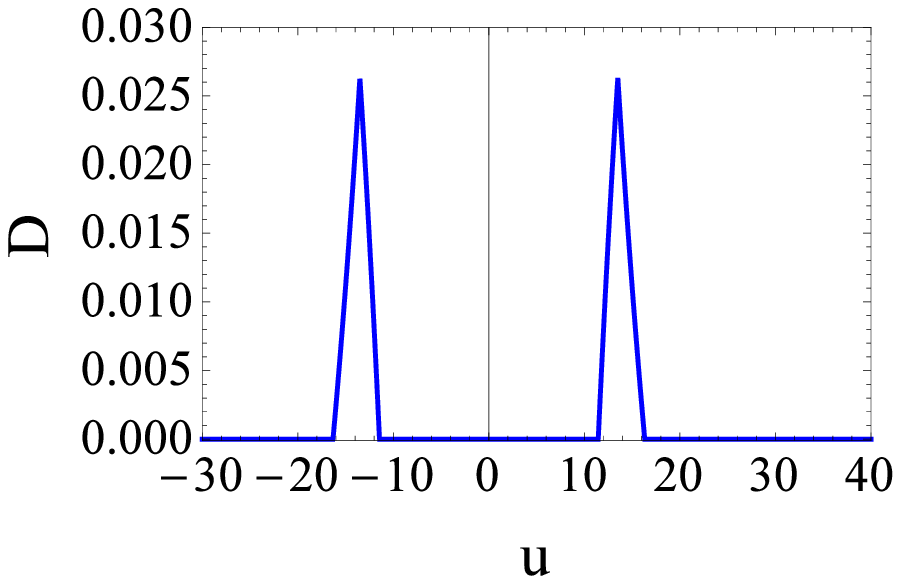}
\caption{The eigenstates (\ref{eigenvalueEq}) of the Hubbard model for three fermions on three sites  show a pinning-nonpinning transition. The distance $D^{(3,6)}(u)$ of the NON $\vec{\lambda}(u)$ to the polytope boundary for the ground state $|\Psi_1(u)\rangle$ (top) is identical to that for the highest exited state $|\Psi_3(u)\rangle$ (not shown) up to a reflection $u\rightarrow -u$.
The excited state $|\Psi_2(u)\rangle$ (bottom) shows a reentrance behavior of pinning in form of four pinning-nonpinning transition.}
\label{fig:saturationBD}
\end{figure}
Indeed, as shown on the upper side of Fig.~\ref{fig:saturationBD}, \ $D^{(3,6)}(u)$  for the ground state undergoes a pinning-nonpinning transition. $D^{(3,6)}(u)$ for the excited state $|\Psi_2(u)\rangle$ (lower panel of Fig.~\ref{fig:saturationBD}) shows indeed that an expected \emph{reentrance} phenomenon is present, a sequence of pinning-nonpinning transitions.

Due to the distinguished role of the ground state we determine its leading behavior on the right side of the `critical' point $u_0$.
From the concrete form of the ground state $|\Psi_1(u)\rangle$ (see Appendix \ref{app:cubicEigenProb}) one obtains
\begin{equation}
D^{(3,6)}(u)  \simeq  0.026\, (u-u_0) \,\,,\,\,\ \mbox{for}\,u\geq u_0\,.
\end{equation}

\begin{figure}[]
\centering
\includegraphics[scale=0.29]{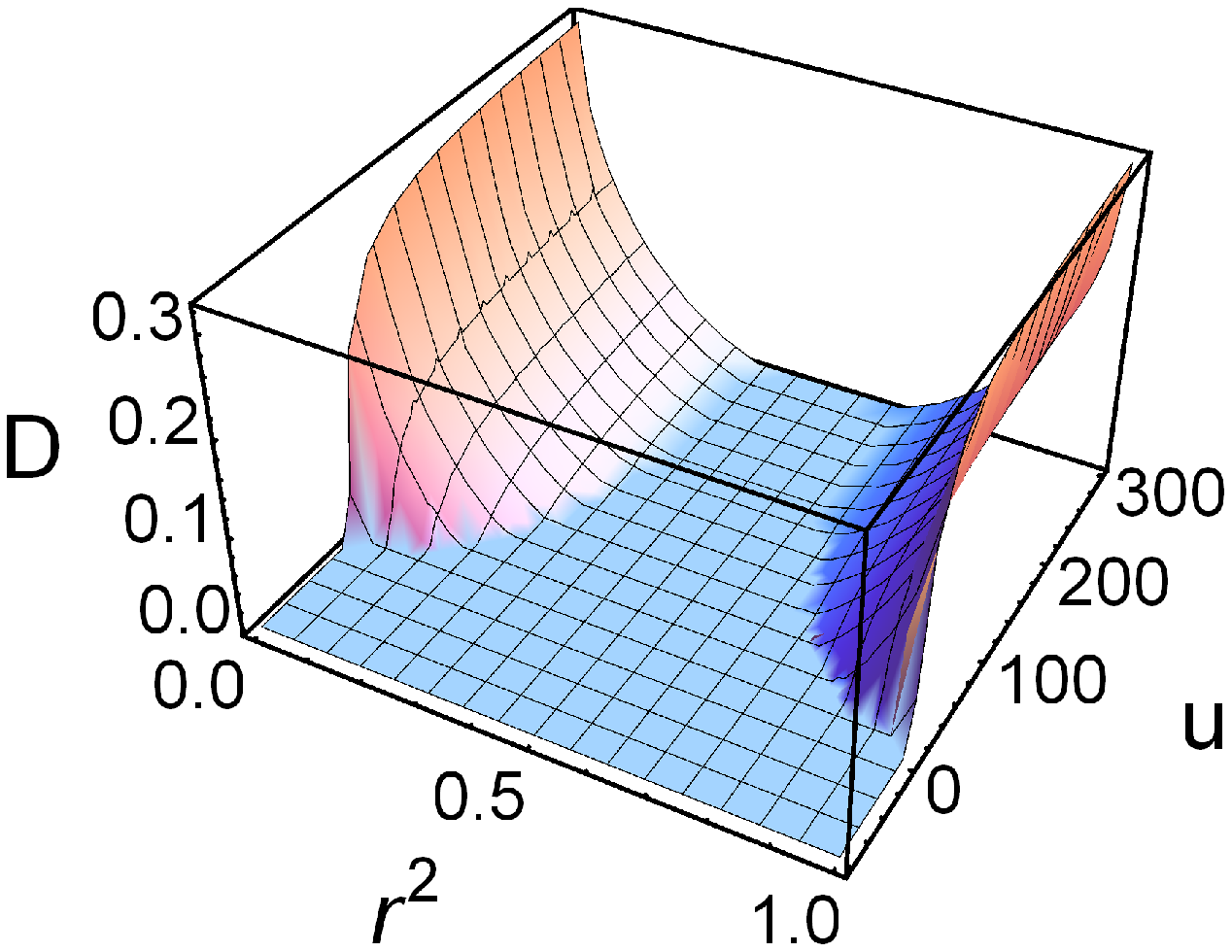}\hspace{0.1cm}
\includegraphics[scale=0.29]{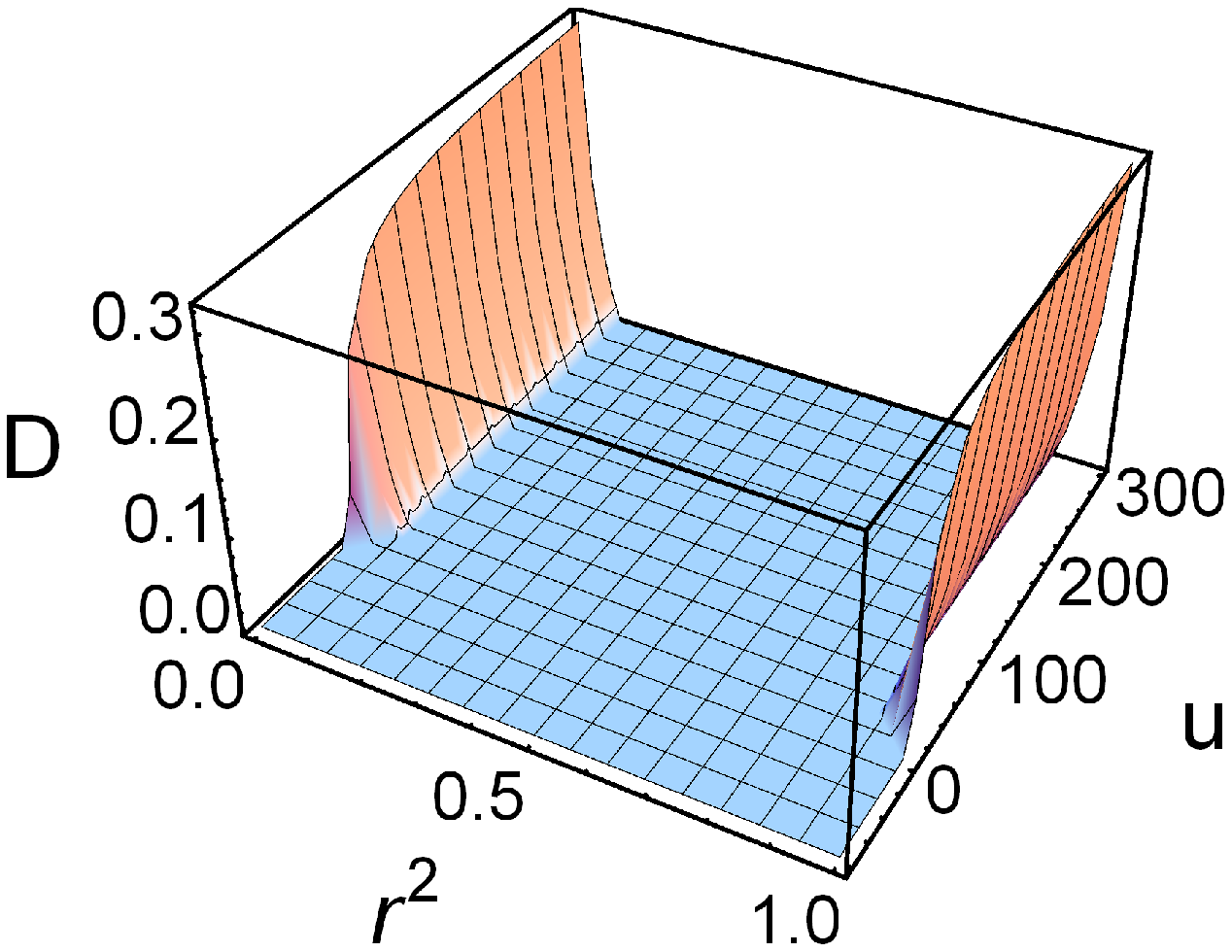}
\caption{Saturation $D(u)$ for the ground state $|\Psi(u)\rangle = \sqrt{1-r^2} |\Psi_1(u)\rangle + r\,e^{i \varphi}\mathcal{I}_{Q}|\Psi_1(u)\rangle $ of the Hubbard model with three sites and three fermions for the two extremal cases $\varphi= 0$ (left) and $\varphi= \frac{\pi}{3}$ (right).}
\label{fig:saturationBDsuperU}
\end{figure}

So far we have restricted ourselves to symmetry-adapted eigenstates. However, since experimental realizations of the Hubbard model typically prepare states with fixed spin quantum numbers (see e.g.~\cite{JochimHubb}) the most general ground state takes the form
\begin{equation}\label{gsHubbard2}
|\Psi_{\xi,\zeta}(u)\rangle \equiv \xi |\Psi_1(u) \rangle + \zeta \mathcal{I}_{Q}|\Psi_1(u)\rangle\,,
\end{equation}
but has still well-defined spin quantum numbers $S=\frac{1}{2}$, $M=\pm\frac{1}{2}$.
Here $|\xi|^2+|\zeta|^2=1$, $\mathcal{I}_{Q}$ is the $3$-fermion inversion operator for the reciprocal ($Q$) lattice and $|\Psi_1(u) \rangle$ has the form (\ref{gsHubbard1}) where the coefficients follow from the eigenvalue problem (\ref{eigenvalueQ}) and (\ref{eigenvalueEq}).
We set $\zeta\equiv |\zeta| e^{i\varphi}$ and without loss of generality we choose $\xi= |\xi| = \sqrt{1-|\zeta|^2}$. Due to a $\frac{2\pi}{3}$-symmetry of the NON (see Appendix \ref{app:cubicsuper}) we present the pinning results in Fig.~\ref{fig:saturationBDsuperU} for the two extremal cases $\varphi=0, \frac{\pi}{3}$.
There we can see that pinning is uniform in $u$  whenever $|\zeta|^2 = |\xi|^2 =r^2= \frac{1}{2}$. For all other superpositions it vanishes for some critical value $u_0(\zeta)$. Moreover, since this critical value $u_0(\zeta)$ is minimal for either $\zeta=0$ or $|\zeta| =1$, we conclude that superposing $|\Psi_1(u)\rangle$ and $\mathcal{I}_{Q}|\Psi_1(u)\rangle$ extends the $u$-regime of pinning from $(-\infty,12.86]$ to some larger interval $(-\infty, u_0(\zeta)]$, $u_0(\zeta) \geq 12.86$. Furthermore, as already suggested by Fig.~\ref{fig:saturationBDsuperU} it turns out that the relative phase $\varphi=\frac{\pi}{3}$ favors pinning best.

\section{Larger settings}\label{sec:larger}
In this section we study the next larger systems. However, since complete families of GPC are not known yet for cases
corresponding to more than $N=5$ fermions and more than $r=5$ lattice sites we restrict to $N\leq 5$ and $r\leq 5$. The GPC for all those cases
can be found in \cite{Altun}. We diagonalize the corresponding Hamiltonians exactly by numerical methods.
In the following we present as representative cases the two cases $(N,r)=(3,4),(5,5)$. The other cases $(3,5),(4,4),(4,5)$ are qualitatively similar. Note also, that due to the particle-hole symmetry \cite{Fradkin}, the NON for the cases $(N,r)$ with $N>r$ follow from those for $N<r$.\\

\subsection{Three fermions on four sites}
\begin{center}
\begin{figure}[h]
\centering
\includegraphics[scale=0.29]{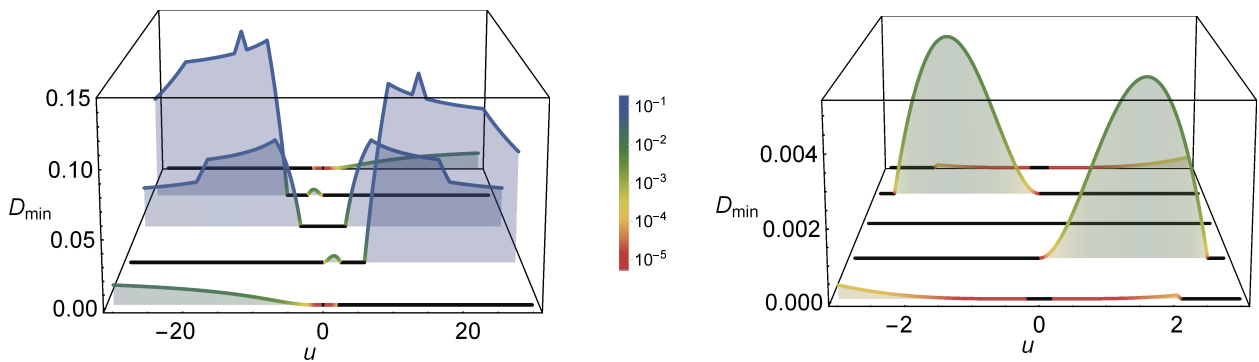}
\caption{For the Hubbard model with three fermions on four sites all five eigenstates with quantum numbers $(S,M,K)=(\frac{1}{2},\frac{1}{2},1)$ are explored w.r.t.~possible (quasi)pinning. The curves $D_{min}(u)$ showing the minimal distance of the NON $\vec{\lambda}(u)$ to the polytope boundary are ordered according to increasing energy from the front to the back. Pinning is shown in black and the strength of quasipinning is visualized in form of different colors.}
\label{fig:g34}
\end{figure}
\end{center}

\begin{figure}[]
\centering
\includegraphics[scale=0.48]{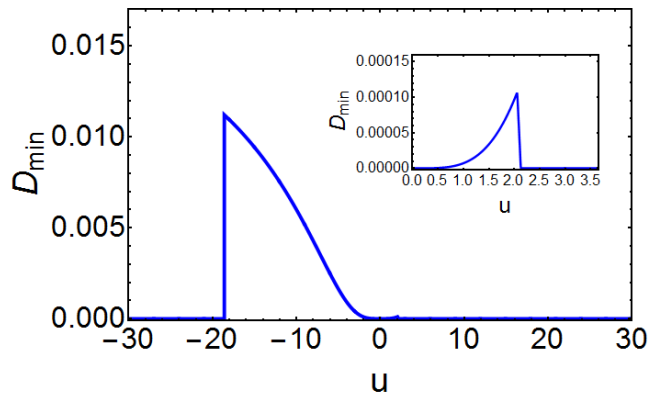}
\caption{For the ground state of the Hubbard model with three fermions on four sites possible (quasi)pinning is explored. For very negative on-site interactions $u$ the ground states exhibits always pinning which vanishes at a `critical' point $u_1\approx -18.6$ in form of a sharp pinning-nonpinning transition. Also quasipinning-pinning transitions can be found, namely at the `critical' points $u_2=0$ and $u_3\approx 2.1$ (see inset).}
\label{fig:Pin34}
\end{figure}

For the Hubbard model with three spin-$\frac{1}{2}$ fermions on four lattice sites we explore possible pinning for several symmetry-adapted eigenstates. In Fig.~\ref{fig:g34} we present the results, i.e.~the minimal distance $D_{min}(u)$ of the vector $\vec{\lambda}(u)$ of NON to the polytope boundary $\partial\mathcal{P}_{3,4}$, for the $5$-dimensional subspace with quantum numbers $(S,M,K)=(\frac{1}{2},\frac{1}{2},1)$. The eigenstates for the other subspaces show similar behavior. $D_{min}(u)$ for the five eigenstates are ordered from the front to the back according to increasing energy. The first and fifth as well as the second and fourth state behave identically up to a reflection of $u$ at $u=0$ \footnote{This follows from the $U(1)$-gauge transformation $c_{j\sigma} \rightarrow e^{i j \pi} c_{j\sigma}$ applied to Hamiltonian (\ref{HamHubbard}) which transforms $H(t,U)$ for the case of for an even number of sites into $-H(t,-U)$.}.
$D_{min}$ for the eigenstate with minimal energy shows transitions between pinning (shown in black) and quasipinning, which is here defined as $D_{min}\leq 10^{-3}$. The second eigenstate in this subspace exhibits pinning for negative $u$. For positive $u$ there is a pinning-quasipinning-pinning transition followed by nonpinning for larger $u$. The third eigenstate is pinned in the interval $[-3.75,3.75]$ and nonpinned in the remaining regime.
\onecolumngrid
\twocolumngrid
It is instructive and also relevant for possible experimental realizations of the pinning effect to study the global ground state separately. For the $u$-interval $(-18.6,18.6)$ the ground state belongs to the blocks $(S,M,K)=(\frac{1}{2},\pm\frac{1}{2},\pm1)$ which were already discussed and presented in Fig.~\ref{fig:g34}. At `critical' points $u\approx \pm 18.6$ energy eigencurves cross and the ground state changes its quantum numbers to $(S,M,K)=(\frac{1}{2},\pm\frac{1}{2},\pm 2)$ for $u<-18.6$ and $(S,M,K)=(\frac{3}{2},\pm\frac{3}{2}, 0), (\frac{3}{2},\pm\frac{1}{2}, 0)$ for $u>18.6$. The ground state pinning behavior is shown in Fig.~\ref{fig:Pin34}. The behavior in the regime $(-18.6,18.6)$ is already known from Fig.~\ref{fig:g34} and for $|u|>18.6$ there is always pinning. The nonanalytic behavior of the distance $D_{min}(u)$ of the NON $\vec{\lambda}(u)$ to the polytope boundary at `critical' points has different origins.
First, recall that the Hamiltonian (\ref{HamHubbard}) depends analytically on the on-site interaction $u$ and the energy spectrum and eigenstates inherit this analyticity. However, the ordering of the eigenstates according to increasing energies may violate this analyticity.
\begin{figure}[]
\centering
\includegraphics[scale=0.126]{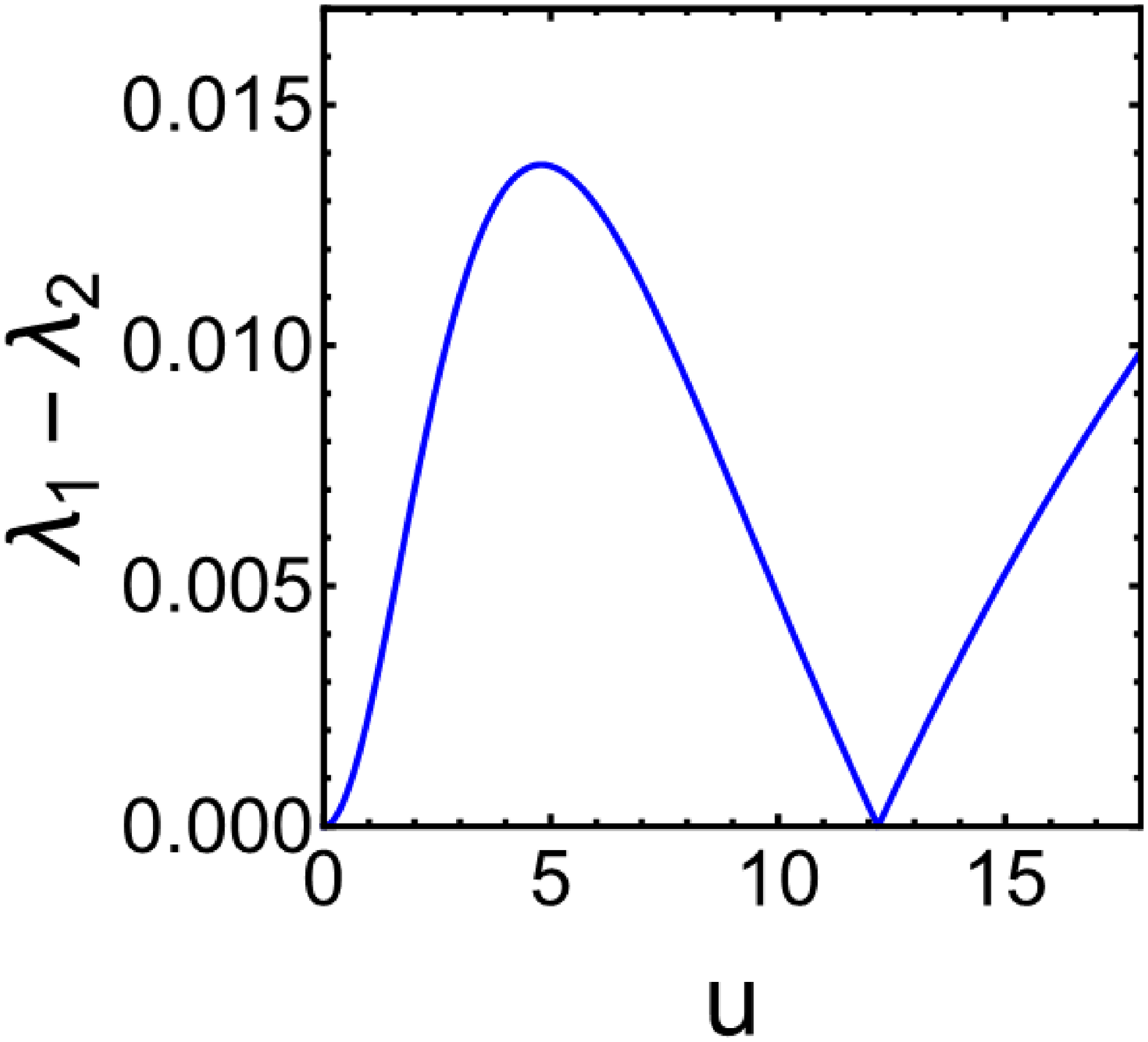}\hspace{0.1cm}
\includegraphics[scale=0.13]{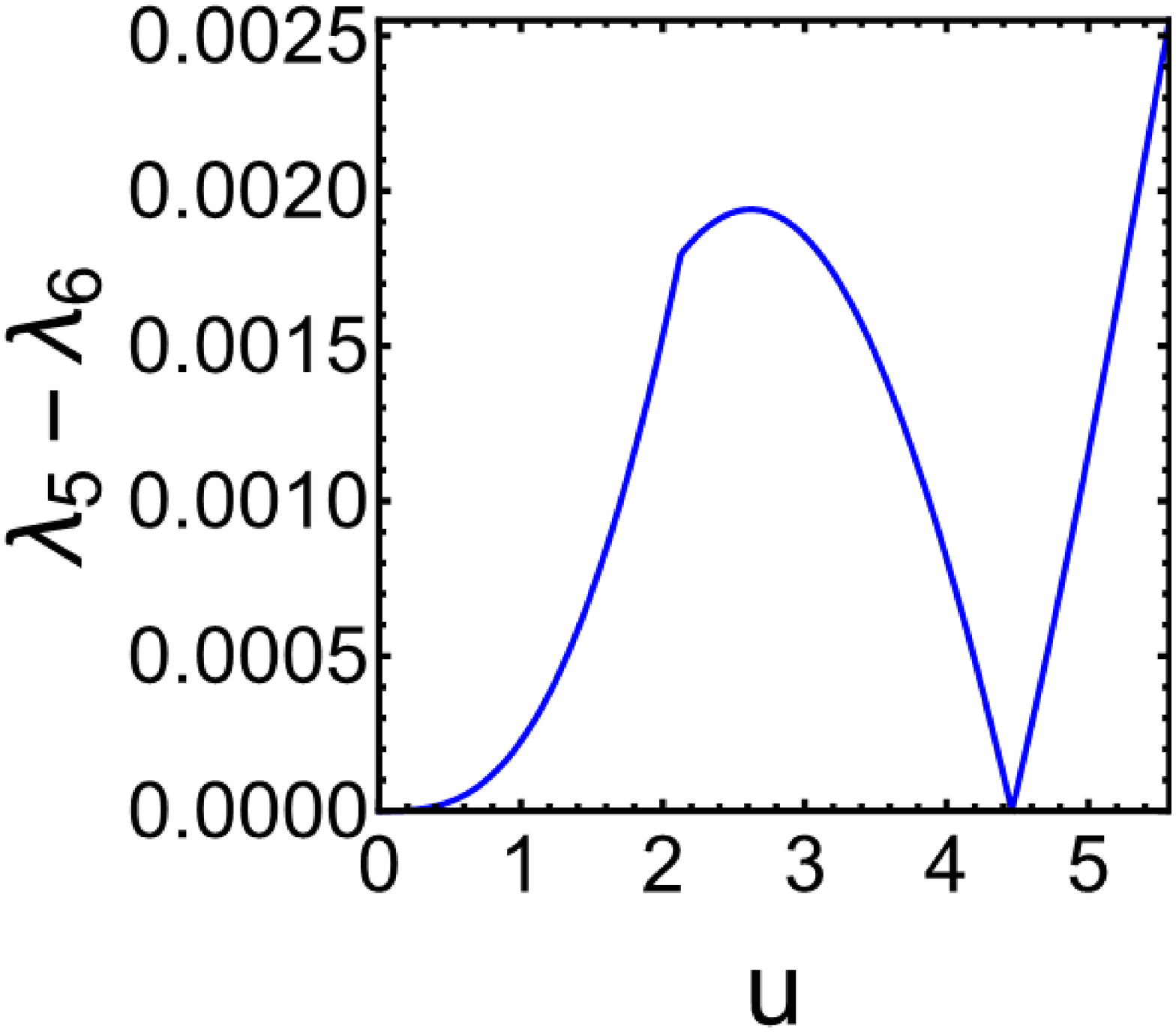}\hspace{0.1cm}
\includegraphics[scale=0.135]{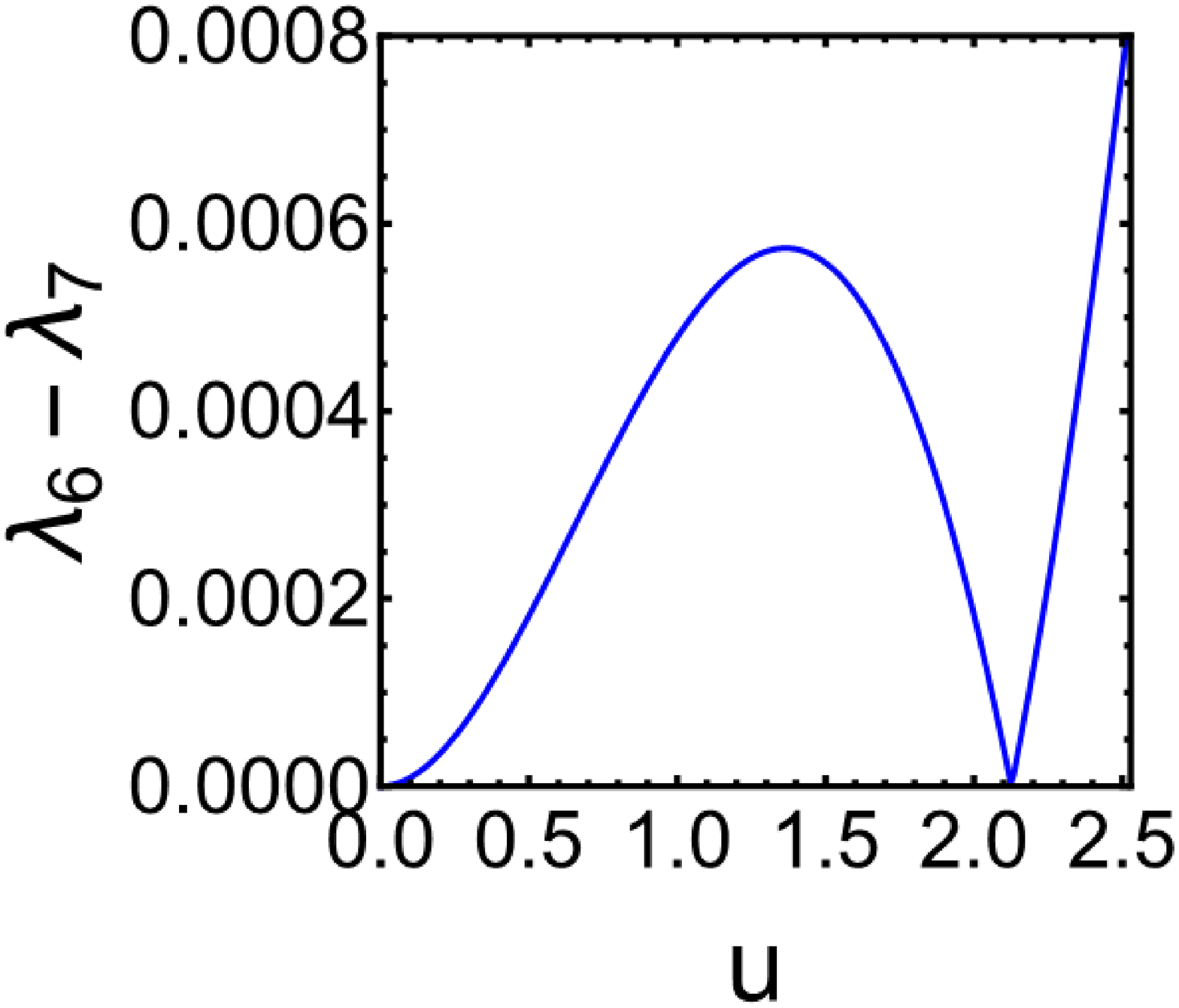}
\caption{For the ground state of the Hubbard model with three fermions on four sites only the distances $\lambda_{i}(u)-\lambda_{i+1}$ for $i=1,5,6$ of decreasingly ordered  NON $\lambda_i(u)$ have nonanalytic points. Such nonanalytic behavior for some `critical' values for $u$ can change the underlying pinning behavior $D_{min}(u)$ of the ground state as shown in Fig.~\ref{fig:Pin34}}
\label{fig:NONdif}
\end{figure}
This happens as already mentioned above at the `critical' point $u=-18.6$. At the other `critical' point $u=+18.6$ $D_{min}(u)$ is analytic since both quantum states whose energy curves are crossing show identical pinning behavior (both are pinned). Even for a given analytic $N$-fermion state $|\Psi(u)\rangle$ nonanalytic behavior of $D_{min}(u)$ is possible. Although $\hat{\rho}_1$ (recall (\ref{1RDO})) and therefore also its spectrum (NON) and natural orbitals inherit this analyticity the ordering of the NON $\lambda_{i}(u)\geq \lambda_{i+1}(u)$ may violate it and could therefore lead to `critical' points for $D_{min}(u)$. This happens at $u\approx 2.1$ (see inset of Fig.~\ref{fig:Pin34}). To verify this we study the distances of successive NON and present those three pairs in Fig.~\ref{fig:NONdif} which indicate crossings of NON. There is indeed a crossing at $u\approx 2.1$ since $\lambda_6-\lambda_7$ vanishes there. But why do the crossings at $u=4.4, 12.1$ indicated in the first two graphics in Fig.~\ref{fig:NONdif} not lead to `critical' points for $D_{min}(u)$? The reason is obvious. Crossing of NON has only an influence on $D_{min}(u)$ if the corresponding two NON do enter $D_{min}$ differently (recall the form (\ref{GPC})), i.e.~with different coefficients $\kappa_i$. To verify this for the ground state studied in this section we first need to know the polytope facet which is closest to $\vec{\lambda}(u)$, i.e.~we need to find the most saturated GPC (\ref{GPC}). For $u\geq 2.0$ this is the constraint
\begin{equation}\label{GPC34a}
D(\vec{\lambda}) = 2-(\lambda_1+\lambda_2+\lambda_4+\lambda_7)\geq 0.
\end{equation}
Indeed, the NON $\lambda_6, \lambda_7$ have different weights, $\kappa_6=0$ and $\kappa_7=-1$. This is not true for the other pairs of NON, $\lambda_1,\lambda_2$ and $\lambda_5,\lambda_6$ since $\kappa_1=\kappa_2=-1$ and $\kappa_5=\kappa_6=0$. Consequently, the pinning behavior is also analytic around $u\approx 4.4, 12.1$.
A third reason for nonanalytic behavior of $D_{min}(u)$ is given when the GPC which is most saturated changes, i.e.~we have a crossing of two saturations $D(u), D'(u)$. This happens at the point $u\approx 2.0$ (see sharp kink in the inset of Fig.~\ref{fig:Pin34}) where the GPC (\ref{GPC34a}) and $D'(\vec{\lambda})= 1-\lambda_1-\lambda_2+\lambda_3\geq0$ interchange their role as most saturated GPC.

\subsection{Five fermions on five sites}
\begin{figure}[]
\includegraphics[scale=0.27]{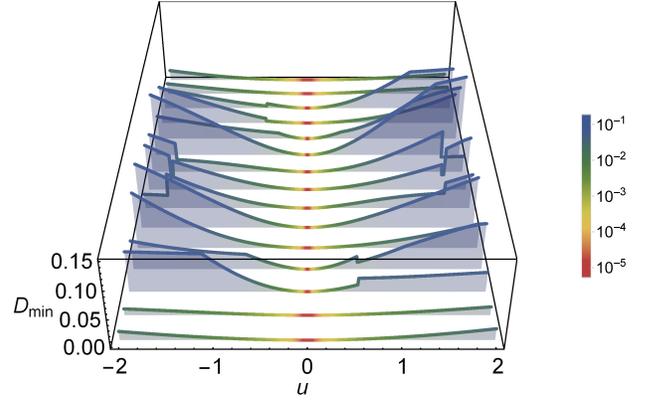}
\caption{For the Hubbard model with five fermions on five sites all fifteen eigenstates with quantum numbers $(S,M,K)=(\frac{1}{2},\frac{1}{2},1)$ are explored w.r.t.~possible (quasi)pinning. The curves $D_{min}(u)$ showing the minimal distance of the NON $\vec{\lambda}(u)$ to the polytope boundary are ordered according to increasing energy from the front to the back. Pinning is given (for each eigenstate) only for the single point $u=0$ (not visible in the graphic) and the strength of quasipinning is visualized in form of different colors. Quasipinning is found for several eigenstates in the neighborhood of $u=0$.}
\label{fig:g55}
\end{figure}

As a third case we study five fermions on five lattice sites. The ground state (with multiplicity four) belongs to the subspaces $(S,M,K)=(\frac{1}{2},\pm\frac{1}{2},\pm 1)$. Therefore, we explore possible (quasi)pinning for the eigenstates with $(S,M,K)=(\frac{1}{2},\frac{1}{2},1)$. The results are shown in Fig.~\ref{fig:g55}. The eigenstates are ordered from the front to the back according to increasing energy. In particular, the first curve $D_{min}(u)$ describes the absolute ground state.  For several eigenstates pinning of NON $\vec{\lambda}$ occurs only for the isolated point $u=0$. In the neighborhood of $u=0$ we find quasipinning which eventually changes to nonpinning for $|u| \gtrsim 0.5$.
The eigenstates of the other subspaces with $S=\frac{1}{2}$ show very similar behavior as those for $K=1$ shown in Fig.~\ref{fig:g55}. For $S=\frac{3}{2},\frac{5}{2}$ the corresponding subspaces are much smaller and the pinning behavior shows more transitions similar to the results for $S=\frac{1}{2}$ for three fermions on four sites presented in Fig.~\ref{fig:g34}.

Finally, it should be also mentioned that linearly superposing degenerate eigenstates belonging to subspaces with different quantum numbers leads again to enhancement of pinning as discussed at the end of Sec.~\ref{sec:pinning}.

\section{Role of symmetries and structural implications of pinning}\label{sec:struct}
In this section we emphasize the strong relation between symmetries and pinning.
One of the most remarkable consequences of pinning as an effect in the $1$-particle picture is that it allows to reconstruct the structure of the
corresponding $N$-fermion quantum state $\ket{\Psi}$. In addition, $\ket{\Psi}$ is significantly simplified (see e.g.~\cite{Kly1,CSthesis,CSQMath12,CSQuasipinning}). To explain this, notice that for NON saturating all Pauli exclusion principle constraints,
$\ket{\Psi}$ can be written as a single Slater determinant,
\begin{equation}\label{Eq:PEPstruct}
\vec{\lambda}= \vec{\lambda}_{HF}\equiv(\underbrace{1,\ldots,1}_N,\underbrace{0,\ldots,0}_{d-N})\,\,\Rightarrow \,\, \ket{\Psi}=\ket{i_1,\ldots,i_N}\,.
\end{equation}
This structural simplification for pinning to the Hartree-Fock point $\vec{\lambda}_{HF}$, which is a vertex of the polytope $\mathcal{P}_{N,d}$, generalizes to arbitrary points on the polytope boundary: By expanding $|\Psi\rangle$ in Slater determinants built up from its own natural orbitals $|i\rangle$,
\begin{equation}\label{Eq:GPCstruct}
|\Psi\rangle = \sum_{1\leq i_1<\ldots<i_N\leq d}\,c_{i_1,\ldots,i_N}\,|i_1,\ldots,i_N\rangle\,,
\end{equation}
many of the expansion coefficients $c_{i_1,\ldots,i_N}$ need to vanish rigorously in case of pinning.
This selection rule of Slater determinant for pinning $D(\vec{\lambda})=0$ reads (see e.g.~\cite{CSQuasipinning})
\begin{equation}\label{eq:SelRule}
\hat{D}_{\Psi}\ket{\bd{i}} \neq 0 \quad \Rightarrow \quad c_{\bd{i}}=0\,.
\end{equation}
Here $\ket{\bd{i}} \equiv \ket{i_1,\ldots,i_N}$ and
\begin{equation}\label{eq:Dop}
\hat{D}_{\Psi}\equiv D(\hat{n}_1,\ldots,\hat{n}_d)\,,
\end{equation}
where $\hat{n}_j$ is the particle number operator for the $j$-th natural orbital of $\ket{\Psi}$. Since $D(\cdot)$ is linear (recall Eq.~(\ref{GPC})),
$\hat{D}_{\Psi}$ is a (hermitian) $1$-particle operator.

The structural simplifications of $\ket{\Psi}$ by pinning lead to a reduction of entanglement described, e.g., by the von Neumann entropy of the $1$-particle reduced density operator $\hat{\rho}_1$. This follows from the fact that the maximal entanglement corresponds to NON $\vec{\lambda}=(\frac{N}{d},\ldots,\frac{N}{d})$, which never saturate any GPC. It is a future challenge to derive optimal upper bounds on the entanglement entropy given pinning. Furthermore, since excitations depend qualitatively on the structure of the ground state, pinning may also influence the low temperature properties.

For systems of interacting fermions one does not expect any of the coefficients $c_{\bd{i}}$ of the superposition (\ref{Eq:GPCstruct}) to vanish.
Consequently, pinning seems unlikely or even impossible.
However, as the analysis of the Hubbard model has shown, $1$-particle symmetries, like the translational or spin symmetry, cause such a significant reduction of Slater determinants (see e.g.~(\ref{gsHubbard1})) and eventually can cause pinning. Yet, our results also show that the presence of symmetries does not automatically imply pinning.

Besides the main insight that symmetries of the Hamiltonian favor pinning, a converse statement is also possible.
According to Eqs.~(\ref{Eq:GPCstruct}),(\ref{eq:SelRule}) and (\ref{eq:Dop}), the occurrence of pinning reveals a $1$-particle symmetry of the corresponding quantum state $\ket{\Psi}$. It is generated by the `observable' $\hat{D}_{\Psi}$ and reads
\begin{equation}\label{eq:Symnew}
e^{-i\varphi \hat{D}_{\Psi}}\ket{\Psi}= \ket{\Psi}\quad,\,\forall\varphi\,.
\end{equation}

All consequences of pinning presented in this section hold approximately for quasipinning, as well \cite{CSQuasipinning}.

\section{Experimental realization of pinning}\label{sec:exp}
In this section we outline two conceptually different ideas for realizing and verifying pinning.
Both of them are based on very recent progress in the field of ultracold fermionic gases simulating the few-site Hubbard model
with full control over its quantum state\cite{JochimHubb}. By focusing laser beams with a high-resolution objective, a potential landscape with a few wells is generated which can be loaded with ${}^6\mbox{Li}$ atoms. This system then simulates the few-site Hubbard model. By independently controlling the intensity and position of the laser beams with an acousto-optic deflector the hopping between the wells can be tuned\cite{JochimHubb}.
Hence, by preparing, e.g., three ultracold fermionic atoms in an optical potential with four local minima located at the vertices of a square, the results found in our work can be checked. This only requires a measurement of the $1$-particle occupation numbers as performed in \cite{JochimHubb} \footnote{In \cite{JochimHubb} the occupancies of the lattice sites were measured. It is not clear to us whether their techniques also allow to measure the occupancies in the momentum representation. For bosons such momentum distributions were successfully measured already in 1995 to verify Bose-Einstein condensation \cite{BECKetterle, BECWieman}.}.
\begin{figure}[]
\centering
	\includegraphics[width=7.5cm]{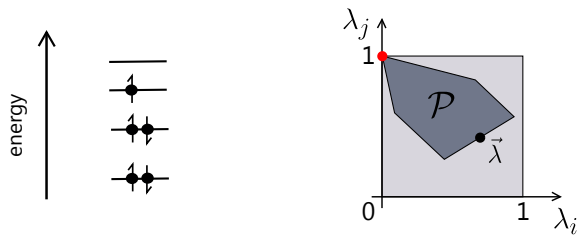}
\caption{On the left side, the Pauli exclusion principle prevents the valence electron from decaying to lower lying energy shells. This kinematical restriction for $\vec{\lambda}=(1,\ldots,1,0,\ldots)\in \partial \mathcal{P}$ (red dot) generalizes to arbitrary `points' $\vec{\lambda}$ on the polytope boundary $\partial \mathcal{P}$ (schematic illustration on the right side).}
	\label{fig:robust1}
\end{figure}

Besides such a direct verification of pinning by measuring NON an indirect and more fault-tolerant approach is also possible. It makes use of the fact that pinning potentially restricts the dynamics of the corresponding system.
This is illustrated in Fig.~\ref{fig:robust1}. On the left side, the valence electron cannot decay to lower lying energy states since those are already occupied. Such a kinematical restriction by Pauli's exclusion principle is generalized by GPC to arbitrary `points' $\vec{\lambda}$ on the polytope boundary. There, as presented on the right side of Fig.~\ref{fig:robust1}, pinned $\vec{\lambda}$ can never leave the polytope.
A complete mathematical analysis reveals even stronger implications. The linear response of pinned $\vec{\lambda}$ to a perturbation of the corresponding $N$-fermion quantum state $\ket{\Psi}$ is restricted to the polytope facet \cite{Alex}. This means that $\vec{\lambda}$ in leading order cannot leave the polytope facet under any small perturbation of the system and in particular cannot move to the interior of $\mathcal{P}$. For quasipinning, this holds approximately, as well.
\begin{figure}[]
\centering
	\includegraphics[width=7.5cm]{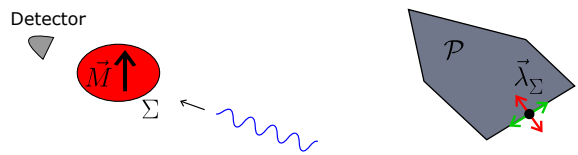}
\caption{Left side: A laser wave (blue) couples to the magnetic moment of a system $\Sigma$. However, since $\Sigma$ is prepared in a pinned state its linear response (right) may be suppressed and would lead to a transparency effect (see text).}
	\label{fig:robust2}
\end{figure}
In Fig.~\ref{fig:robust2}, a set-up is proposed for the indirect experimental verification of pinning by referring to its robustness. A physical system $\Sigma$, e.g., the $3$-fermion Hubbard model, is perturbed by a laser wave (shown in blue) which couples to the magnetic moment $\vec{M}$ of $\Sigma$. By describing the
interaction process from the $1$-particle viewpoint (right side) the linear response of $\Sigma$ may significantly depend, e.g., on the polarization or the direction of the laser wave. In some specific cases (shown in red), the linear response of $\vec{\lambda}$ is expected to be perpendicular to the polytope facet. However, this is in contradiction to the existence of GPC and therefore the interaction between $\Sigma$ and the laser wave has to be suppressed. A detector as shown in the set-up on the left side of Fig.~\ref{fig:robust2} can then detect this non-scattered laser wave.
Furthermore, if $\Sigma$ is given by the few-site Hubbard model, changing its on-site interaction would allow to verify the pinning-nonpinning transitions found in our work.

By considering not only a single few-site Hubbard model but macroscopically many, a \emph{GPC-induced transparency}-effect for that material/gas could be observed. All laser waves with a specific polarization, frequency, etc.~are not scattered and the material is not susceptible to them.

\section{Summary and conclusion}\label{sec:concl}
The fermionic exchange statistics does not only imply Pauli's exclusion principle but leads to even stronger restrictions on fermionic
natural occupation numbers (NON). Therefore, these so-called \emph{generalized Pauli constraints} (GPC) have additional  influence on the behavior and the properties of many-fermion systems. A particular feature is their saturation, i.e.~the \emph{pinning} of NON to the boundary of the allowed region. We have argued
that this pinning effect is very rare, or even impossible,  and that instead \emph{quasipinning} occurs \cite{Kly1,BenavLiQuasi,Mazz14,CSthesis,MazzOpen,CSQuasipinning,BenavQuasi2,RDMFT,Alex,RDMFT2,CS2013QChem}.

As discussed above pinning has strong implications. For instance, it leads to a significant simplification of the many-fermion quantum state $\ket{\Psi}$, accompanied by a reduction of its entanglement. Due to these strong implications of pinning it is highly desirable to find models for which pinning really occurs and which can be realized experimentally.

In form of the few-site Hubbard model we succeeded in providing a first system satisfying these requirements.
For all cases $(N,r)$ with $N\leq5$ fermions on $r \leq 5$ sites we found for many eigenstates finite or even semi-infinite intervals
for the relative onsite-interaction $u$, where their NON $\vec{\lambda}(u)$ are pinned. The end points of those intervals define `critical' points where sharp pinning-nonpinning or pinning-quasipinning transitions are found. As an `order parameter' the minimal distance $D_{min}(u)$ of $\vec{\lambda}(u)$ to the polytope boundary $\partial \mathcal{P}$ has been used. We also studied the origin of those transitions which turns out to be either a crossing of eigenenergies or a crossing of NON $\lambda_{i}(u),\lambda_{i+1}(u)$.

The basic reason behind our findings is the presence of rotational (in spin space), and particularly of translational symmetry. This immediately implies that the natural spin orbitals (the eigenstates of the $1$-particle reduced density operator) are the  Bloch states $|k\rangle$ multiplied by a spin state $\ket{\uparrow},\ket{\downarrow}$. Hence the
natural spin orbitals  are known from the very beginning, in contrast to atomic and molecular systems.  As a consequence the NON gain direct physical relevance.

We have suggested two different experimental set-ups for the verification of the pinning-effect, as well as the pinning-to-nonpinning transition. Note that the existence of the nonpinned `phase' proves that symmetry does not automatically implies pinning.
Since GPC emerge from the antisymmetry, the experimental verification of the pinning effect would also provide additional evidence for the antisymmetry of  fermionic wave functions.

The focus of the present contribution has been on pinning. It will be a challenge for future work to investigate its implications. One of them is the simplification of the corresponding many-fermion quantum state.  It will be interesting to study
how far the simplification of   $\ket{\Psi}$ influences its entanglement and the low-lying excitations, responsible for the low temperature behavior.

\paragraph*{Acknowledgements.---}
We would like to thank M.~Christandl, P.G.J.~van Dongen, D.~Gross, D.~Jaksch, A.~Lopes, M.~Rizzi, N.~Schuch, F.~Tennie and J.~Whitfield for helpful discussions. We also gratefully acknowledge financial support from the Swiss National Science Foundation (Grant P2EZP2 152190) and the Oxford Martin Programme on Bio-Inspired Quantum Technologies.

\appendix
\section{Diagonalization of the subspace Hamiltonian (\ref{eigenvalueQ})}\label{app:cubicEigenProb}
In this section we solve the eigenvalue problems (\ref{eigenvalueEq}) where the Hamiltonian is given by Eq.~(\ref{eigenvalueQ}).

The eigenvalues are the roots of the characteristic polynomial $P_u(E)$ of (\ref{eigenvalueQ}), i.e.~we need to solve the cubic equation
\begin{equation}\label{charpoly}
P_u(E)= 6 u+ E \left(u^2-9\right)-2 E^2 u+E^3 =0\,.
\end{equation}
Since Appendix \ref{app:cubicsuper} requires details on the structure of the roots of cubic equations we present in this section a few more details (see e.g.~also \cite{abramo}) than necessary . For the canonic form
\begin{equation}\label{cubicform}
x^3+ax^2+bx+c=0
\end{equation}
we first consider the quantities $Q$ and $R$, defined as
\begin{equation}\label{cubicParQ}
Q \equiv \frac{a^2-3b}{9}\qquad,\,\,  R\equiv\frac{2 a^3-9ab+27c}{54}\,.
\end{equation}
By defining
\begin{equation}\label{cubicParTh}
\Theta \equiv \arccos{\left(R/\sqrt{Q^3}\right)}
\end{equation}
the three real roots of (\ref{cubicform}) follow as
\begin{equation}\label{roots}
x_k = -2 \sqrt{Q}\,\cos{\left(\frac{\Theta +2\pi k}{3}\right)} -\frac{a}{3}\,,\,k=0,1,2\,.
\end{equation}

For our concrete characteristic polynomial (\ref{charpoly}) we have
\begin{equation}
Q = \frac{u^2}{9}+3\qquad,\,\,  R = \frac{u^3}{27}\,.
\end{equation}
and as real roots (energy eigenvalues) $E_1(u),E_2(u)$ and $E_3(u)$ we find
\begin{widetext}
\begin{eqnarray}
E_k &=& \frac{2}{3} \left\{u-\sqrt{u^2+27} \cos \left[\frac{1}{3} \arccos\left(u^3/\sqrt{\left(u^2+27\right)^3}\right)+\frac{2\pi (1-k)}{3}\right]\right\}\nonumber\,.
\end{eqnarray}
\end{widetext}
Here, we ordered the eigenvalues differently than in (\ref{roots}) to be consistent with the labeling in Fig.~\ref{fig:energyplots}.
Notice also the symmetries of these energies,
\begin{equation}\label{energysymetries}
E_1(-u)= -E_3(u)\quad \mbox{and}\quad E_2(-u)= -E_2(u)\,.
\end{equation}

The corresponding unnormalized eigenvectors $\vec{v}_j(u) \equiv (\tilde{\alpha}_j(u),\tilde{\beta}_j(u),\tilde{\gamma}_j(u)), j=1,2,3$ of matrix (\ref{eigenvalueQ}) follow directly from (\ref{eigenvalueQ}). We find
\begin{eqnarray}
\tilde{\alpha}_j(u) &=& u+3-E_j(u) \nonumber \\
\tilde{\beta}_j(u) &=& u -3-E_j(u) \nonumber \\
\tilde{\gamma}_j(u) &=&u-4 E_j(u)+\frac{3}{u} \left[E_j(u)^2-9\right]\,.
\end{eqnarray}
which after division by $\sqrt{|\tilde{\alpha}_j(u)|^2+|\tilde{\beta}_j(u)|^2+|\tilde{\gamma}_j(u)|^2}$ yields the normalized coefficients of the eigenstates (\ref{gsHubbard1}).

\section{Calculation of the NON for the superposition (\ref{gsHubbard2})}\label{app:cubicsuper}
In this appendix we calculate the NON for the more general ground state (\ref{gsHubbard2}). Since (\ref{gsHubbard2}) is not symmetry-adapted to the lattice translation anymore, the corresponding $1$-particle reduced density operator $\hat{\rho}_1$ represented w.r.t.~the Bloch states $\{|k,\sigma\rangle\}$ is not diagonal anymore. However, since (\ref{gsHubbard2}) is still an $S_z$-eigenstate, the new natural orbitals are still spin eigenstates w.r.t.~$z$-axis and $\hat{\rho}_1$ splits into blocks according to Eq.~(\ref{1rdoBlocks}). The specific normalizations of both blocks together with the three equalities (\ref{GPC36}) imply again that for each eigenvalue $n_j^{\uparrow}$
of $\hat{\rho}_1^{\uparrow}$ there exists a corresponding eigenvalue $n_i^{\downarrow}$ of $\hat{\rho}_1^{\downarrow}$ such that their sum equals one, $n_j^{\uparrow}+ n_{4-j}^{\downarrow}=1, j=1,2,3$, where the triples $\{n_j^{\uparrow}\}$ and $\{n_j^{\downarrow}\}$ are ordered decreasingly. Hence, it is sufficient for our purpose to diagonalize just the block $\hat{\rho}_1^{\downarrow}$. By representing it w.r.t.~the states $\{|k\downarrow\rangle\}$ we obtain the following density matrix
\begin{equation}\label{rho1down}
 \hat{\rho}_1^{\downarrow} = \left(\begin{array}{ccc}|
 \alpha|^2& - \xi^\ast \zeta \alpha^\ast \gamma& - \xi \zeta^\ast \alpha^\ast \gamma \\
 - \xi\zeta^\ast \alpha \gamma^\ast&|\zeta|^2 |\beta|^2+|\xi|^2 |\gamma|^2& - \xi^\ast \zeta |\beta|^2\\
 -\xi^\ast \zeta \alpha \gamma^\ast& -\xi \zeta^\ast  |\beta|^2& |\xi|^2 |\beta|^2+|\zeta|^2 |\gamma|^2
 \end{array}\right)\,.
\end{equation}
The cubic characteristic polynomial $p_u(\lambda)$ reads
\begin{eqnarray}
p_u(\lambda) &=& \lambda^3 + c_2\,\lambda^2+c_1\,\lambda+c_0 \,=\,0\\
c_2 &=& - 1 \nonumber \\
c_1 &=& |\alpha(u)|^2\,|\beta(u)|^2 +  |\alpha(u)|^2\,|\gamma(u)|^2 + |\beta(u)|^2\,|\gamma(u)|^2\nonumber \\
&&-(2-3 |\gamma(u)|^2)\, |\gamma(u)|^2 \,|\zeta|^2\,|\xi|^2 \nonumber \\
c_0 &=& - |\alpha(u)|^2\,|\beta(u)|^2\,|\gamma(u)|^2\,\nonumber \\
&&\times\left(1-3|\zeta|^2\,|\xi|^2 - \zeta^3 \,(\xi^*)^3-  (\zeta^*)^3 \,\xi^3\right)\,.\nonumber
\end{eqnarray}
Its roots follow immediately from Eqs.~(\ref{cubicform}), (\ref{cubicParQ}), (\ref{cubicParTh}) and (\ref{roots}) in Appendix \ref{app:cubicEigenProb}.

As a consistency check, notice also that according to Eq.~(\ref{roots}) we have in general
\begin{equation}
x_1 + x_2 + x_3 = - c_2 \,
\end{equation}
which takes here the value $-c_2=1 = \mbox{Tr}[\hat{\rho}_1^{\downarrow}]$.

Finally, we comment on $\xi,\zeta$-symmetries of $\hat{\rho}_1^{\downarrow}$.
First, note that the vector of NON fulfills
\begin{equation}\label{exchange}
\vec{\lambda}(u;\zeta,\xi) = \vec{\lambda}(u;\xi,\zeta)\,,
\end{equation}
which follows directly from the invariance of (\ref{HamHubbard}) w.r.t.~inversions in the Brillouin zone and the form (\ref{gsHubbard2}).
Moreover, blocks $\hat{\rho}_1^{\downarrow}(u;\xi,\zeta)$ fulfill
\begin{equation}
\hat{\rho}_1^{\downarrow}(u;\xi,\zeta e^{\frac{2 \pi}{3}i}) = U^\dagger\, \hat{\rho}_1^{\downarrow}(u;\xi,\zeta) \,U
\end{equation}
with the unitary matrix $U = \mbox{diag}(1,e^{-\frac{2 \pi}{3}i},e^{\frac{2 \pi}{3}i})$. The same actually also holds for the second block, $\hat{\rho}_1^{\uparrow}$, which is not stated here.
Thus,
\begin{equation}
\vec{\lambda}(u;\xi,\zeta e^{\frac{2 \pi}{3}i}) = \vec{\lambda}(u;\xi,\zeta)\,.
\end{equation}

\bibliography{bibliography2}

\end{document}